\documentclass[12pt]{article}

\usepackage{a4wide}
\usepackage{amsmath}
\usepackage{amsthm}
\usepackage{amsfonts}
\usepackage{amssymb}

\usepackage{cite}

\DeclareSymbolFont{AMSb}{U}{msb}{m}{n}
\DeclareSymbolFontAlphabet{\Bbb}{AMSb}

\newcommand{\Z}{\Bbb{Z}}
\newcommand{\R}{\Bbb{R}}

\theoremstyle{plain}
\newtheorem{theorem}{Theorem}

\newtheorem{proposition}[theorem]{Proposition}

\theoremstyle{plain}

\begin{document}

\thispagestyle{empty}

\vspace*{-80pt} 
{\hspace*{\fill} Preprint-KUL-TF-2000/12} 
\vspace{80pt}

\begin{center} 
{\LARGE Interferencing in coupled Bose-Einstein condensates}\\[25pt]

{\large  
    T.~Michoel\footnote{Research Assistant of the Fund for Scientific Research - Flanders
		(Belgium) (F.W.O.)}\footnotetext{Email: {\tt tom.michoel@fys.kuleuven.ac.be}},
	A.~Verbeure\footnote{Email: {\tt andre.verbeure@fys.kuleuven.ac.be}}
    } \\[25pt]   
{Instituut voor Theoretische Fysica} \\  
{Katholieke Universiteit Leuven} \\  
{Celestijnenlaan 200D} \\  
{B-3001 Leuven, Belgium}\\[25pt]
{March 31, 2000}\\[60pt]
\end{center}

\begin{abstract}\noindent
We consider an exactly soluble model of two Bo\-se\--Ein\-stein condensates with  a
Jo\-seph\-son-type of coupling. Its equilibrium states are explicitly found showing
condensation and spontaneously broken gauge symmetry. It is proved that the total number
and total phase fluctuation operators, as well as the relative number and relative current
fluctuation operators form both a quantum canonical pair. The exact relation between the
relative current and phase fluctuation operators is established. Also the dynamics of
these operators is solved showing the collapse and revival phenomenon.
\\[15pt]
\begin{center}
{\bf Keywords}
\end{center}
Bose-Einstein condensation, Josephson-like junction, phase fluctuations, collapse and
revivals, interferences
\end{abstract}
\newpage

\section{Introduction}

Since the 1995-observations \cite{andersonm:1995, davis:1995,
bradley:1995} of Bose-Einstein condensation (BEC) in trapped alkali
gases, an intense renewed interest is going on in the research of
the physical properties and the nature of Bose condensed systems.
In particular the interference pattern between two overlappping
condensates has been measured, see e.g. \cite{mewes:1996} and many
other recent experimental settings and results.

In this context of interference, the static and dynamic properties
of the phase of the condensate are of major importance. This has
been the subject of many theoretical studies all over the last
years. As a primordial and old question, the very existence of the
phase and/or the phase operator, comes into the picture again.

One encounters continuous efforts to formulate the phase (operator)
in the standard theory of BEC, which we could call the
Bogoliubov-Hartree-Fock theory \cite{nozieres:1990}, in a system
with a finite number of atoms (see e.g. \cite{gardiner:1997}). One
is constantly assuming that the condensation is occuring into a
coherent state of the lowest energy mode of the system. Such a
state fixes a well defined phase and amplitude, but should in stead
exhibit inevitable fluctuations of both these quantities. Or, one
is fixing the number of atoms in the system, i.e. the condensation
takes place in a number state of Fock space, excluding any atom
number fluctuations. Although these basic theoretical difficulties
are now getting ripe in the minds of many researchers in the field,
all kinds of procedures and tricks are permanently invented to wave
away these difficulties. In this paper we take the point of view
that these questions about the character of the quantum state into
which the condensation occurs, and its major properties, are
nevertheless of major importance. By now it is indeed well known
that a condensate state is neither a `pure Fock', nor a `pure coherent'
state in the strict mathematical sense, nor in the physical sense.

As explained above, `simple' coherent states or Fock states lead to
annoying technical difficulties in order to describe and understand
the essentials of many of the experimental challenging measurements
on BEC which are constantly performed. After all, condensation is
up to now, only clearly defined and generally accepted for
homogeneous systems. Of course, we are aware of different
tentatives to introduce decent thermodynamic limits for trapped
gases. With all this knowledge in mind, we focus our attention
here, not on the situation of BEC in trapped gases, but on the
phenomenon of BEC for homogeneous systems, where one has a well
defined thermodynamic limit, and where the occurence of BEC,
accompanied by a spontaneous $U(1)$ - symmetry breaking
\cite{fannes:1982} is well understood.

Furthermore we take into account that the main entries of the
theory of the Bose condensates and their interference patterns are
the {\em particle number fluctuations} and the {\em phase operator
fluctuations}.

The main question is here, can one define rigorously a phase
operator fluctuation and a particle number operator fluctuation of
the condensate? The answer is proved to be positive. It is based on
the notion of {\em fluctuation operator} which was introduced in a
mathematically rigorous framework some time ago
\cite{goderis:1989b, goderis:1990}. We realise however that these
results did not reach so far the majority of the theoretical
physics community. The aim for introducing the notion of
fluctuation operator, was precisely to study the quantum effects
on the level of the fluctuations. We applied this theory of quantum
fluctuation operators already in order to derive exactly rigorous
results on BEC for the Bogoliubov-Hartree-Fock model
\cite{michoel:1999b}.

In section \ref{s:fluc-op}, we describe these results in a language
approachable for non-mathematics minded readers. The celebrated
phase operator is nothing but the canonical fluctuation operator
conjugate to the number operator fluctuation operator. This section
is not just for warming up, but it should also shed a different
light, than one is used to, on the status of the existence and
meaning of the {\em phase operator}.

In section \ref{s:model}, we study a model of two Bose-Einstein
condensates with a Josephson-like coupling. We look at the static
and thermodynamic properties of this model of two condensates. As
far as we know, the problem of Josephson oscillations between
coupled Bose-Einstein condensates has not yet been considered in a
mathematically rigorous manner. Here we present a solvable model in
a full quantum field theoretical setting and in section 
\ref{s:total-ops} and \ref{s:rel-ops}
we give a full description and analysis of the dynamical equations
of the total and relative number operator fluctuations and the
total and relative phase operator fluctuations. About the dynamics
we find the exact oscillatory time behaviour of all these
fluctuation operators. We also detect the so-called collapse and
revival phenomena.

Our work is, as far as we know, the first rigorous one on this topic
for homogeneous systems, it should also put in a new perspective
much of the discussions which are going on in the large activity
dealing with trapped Bosons and their interference patterns (see
e.g. \cite{wright:1997, wang:1999, villain:1998}).

\section{Number and phase fluctuation operators}\label{s:fluc-op}

In order to fix our ideas and in view of the model we describe in section \ref{s:model}
for the study of the phase interference between two condensates, we present the number
and phase fluctuation operators for the imperfect Bose gas (or mean field Bose gas)
\cite{huang:1967, davies:1972}. We
follow the lines of \cite{michoel:1999b} but it should be clear that its validity is much
larger \cite{michoel:2000}.

The leading idea of this section is to make clear that the up to
now rather `mysterious' phase operator, which everybody uses in
the field, but about which there are doubts on its very existence,
has a firm mathematical definition in an equilibrium condensed
state of a Bose gas. It should be realised that such a condensed
state is neither a coherent, nor a Fock state in the technical narrow
sense. It is defined as the {\em fluctuation operator canonically
adjoint to the number fluctuation operator}. This definition is not
just a formal thing, but it gives a physical interpretation of the
phase operator, different from the existing ones. We are not
repeating here all the mathematics of the definition of the phase
fluctuation operator, which can be found in various papers (see
e.g. \cite{goderis:1989, goderis:1990}). We content ourself here in
making these definitions plausible for the imperfect Bose gas.

Let $\Lambda \subset \R^3$ be the centered cubic box of length $L$,
with periodic boundary conditions.  The Boson creation and
annihilation operators in the one-particle state
$\psi_{L,k}(x)=V^{-1/2}e^{ik.x}$, $x\in\Lambda$, $k\in\Lambda^*=
\frac{2\pi}{L}\Z^\nu$ are given by
\begin{equation*}
a^*_{L,k} = \frac{1}{\sqrt V}\int_\Lambda dx\; a^*(x) e^{ik.x}\quad\text{and}\quad
a_{L,k} = \frac{1}{\sqrt V}\int_\Lambda dx\; a(x) e^{-ik.x},
\end{equation*}
with
\begin{equation*}
[a(x), a^*(y)]=\delta(x-y).
\end{equation*}

The imperfect Bose gas is specified by the local
Hamiltonian $H_L$ \cite{fannes:1980b}:
\begin{equation}\label{eq:ham-mf}
H_L = T_L - \mu_L N_L + \frac{\lambda}{2 V} N_L^2,
\end{equation}
where
\begin{align*}
T_L &= \sum_{k\in \Lambda^*} \epsilon_k a^*_{L,k}a_{L,k}\;,\;\; \epsilon_k =
\frac{|k|^2}{2m}\\
N_L &= \sum_{k\in \Lambda^*}  a^*_{L,k}a_{L,k};
\end{align*}
$\lambda > 0$ measures the strength of the mean field inter particle repulsion.

This model is exactly soluble in the thermodynamic limit
$L\to\infty$, keeping the particle density in the Gibbs state
$\omega_L(\cdot)$ for (\ref{eq:ham-mf}) constant, equal to $\rho$,
i.e. for all L:
\begin{equation*}
\frac{1}{V}\omega_L(N_L)=\rho.
\end{equation*}
It is proved rigorously \cite{fannes:1980b} that for $T<T_c$ or $\rho$ large enough,
the limit state $\omega_\beta(\cdot) =\lim_{L\to\infty}\omega_L(\cdot)$ exists as an
integral over ergodic states ($\omega_\beta^\alpha$, $\alpha\in[0,2\pi]$):
\begin{equation*}
\omega_\beta(\cdot) = \frac{1}{2\pi}\int_0^{2\pi}d\alpha\; \omega_\beta^\alpha(\cdot),
\end{equation*}
with
\begin{equation*}
\omega_\beta^\alpha\bigl( e^{i[a^*(f)+a(f)]} \bigr) = e^{-\frac{1}{2}(f,Kf)+
2i\sqrt{\rho_0}|\hat f(0)|\cos\alpha},
\end{equation*}
where
\begin{equation*}
\widehat{Kf}(k) = \frac{1}{2} \hat f(k)\coth\frac{\beta\epsilon_k}{2}\;,\;\;f\in
L^2(\R^\nu).
\end{equation*}

The spontaneous gauge symmetry breaking accompanying the phase transition is visible in
the states $\omega_\beta^\alpha$ having the property:
\begin{equation*}
\lim_{L\to\infty}\omega_\beta^\alpha\bigl( \frac{a^*_{L,0}}{\sqrt{V}}\bigr) =
\sqrt{\rho_0} e^{i\alpha}.
\end{equation*}
Clearly, $\rho_0$ is the condensate and $\alpha$ is the phase of
the order parameter. One also proves that in the state
$\omega_\beta^\alpha$, one has the operator limit:
\begin{equation*}
\lim_{L\to\infty}\frac{a^*_{L,0}}{\sqrt{V}}=\sqrt{\rho_0} e^{i\alpha}.
\end{equation*}
From now on we limit our attention to one of the ergodic states $\omega_\beta^\alpha$ for
some fixed $\alpha$, and without restriction of generality we take $\alpha=0$, and with a
condensate density $\rho_0\not= 0$. For simplicity, denote the state $\omega_\beta^0$ by
$\omega$.

The state $\omega$ does not have the gauge symmetry. The generator of the gauge symmetry
is the number operator
\begin{equation*}
N_L = \int_\Lambda dx\; a^*(x)a(x)
\end{equation*}
with local number density operator $n(x)=a^*(x)a(x)$. The common choice of order parameter
operator is $V^{-1/2}a^\sharp_{L,0}$, or taking a self-adjoint combination:
\begin{equation*}
O_L = \frac{i}{\sqrt{V}}(a^*_{L,0} - a_{L,0}) = \frac{i}{V}\int_\Lambda dx\;
\bigl( a^*(x) - a(x) \bigr),
\end{equation*}
with local order parameter density operator $o(x)=i\bigl(a^*(x)-a(x)
\bigr)$.

We concentrate now on the $k$-mode fluctuations, with $k\not= 0$, of the local number and
order parameter density operators, i.e. on
\begin{align*}
F_{L,k}(n) &= \frac{1}{\sqrt{V}}\int_\Lambda dx\; \bigl(n(x)-\omega(n(x))\bigr) \cos k.x\\
F_{L,k}(o) &= \frac{1}{\sqrt{V}}\int_\Lambda dx\; \bigl(o(x)-\omega(o(x))\bigr) \cos k.x.
\end{align*}
Remark that for all finite $L$, the quantities $F_{L,k}(n)$ and $F_{L,k}(o)$ are operators
and do represent the fluctuations of the number density and of the order parameter
density.

The first tedious question that is posed, is to characterize the limit operators:
\begin{align*}
F_k(n) &= \lim_{L\to\infty} F_{L,k}(n)\\
F_k(o) &= \lim_{L\to\infty} F_{L,k}(o).
\end{align*}
The details of the proof of these limits can be found in
\cite{michoel:1999b}. Here we just mention that the limits are
taken in the sense of a central limit theorem. The main result is
that the limits $F_k(n)$ and $F_k(o)$ are operators on a well
specified Hilbert space, $\tilde{\mathcal{H}}_k$, generated by a normalised vector
$\tilde\Omega_k$ and vectors $F_k(A_1)\dotsb F_k(A_n)\tilde\Omega_k$, with the $A_i$ local
operators, like e.g. $n(x)$ and $o(x)$, and with arbitrary $n$; the scalar product of
$\tilde{\mathcal{H}}_k$ is given by
\begin{equation}\label{e:scal-prod}
\left(F_k(A_1)\dotsb F_k(A_n)\tilde\Omega_k,F_k(B_1)\dotsb F_k(B_n)\tilde\Omega_k\right)
=\delta_{n,m} \text{Perm}\left\lgroup \left(\tilde\Omega_k,F_k(A_i)F_k(B_j)\tilde\Omega_k
\right)\right\rgroup_{i,j}
\end{equation}
with two-point function given by
\begin{equation}\label{e:two-pt}
\left(\tilde\Omega_k,F_k(A_i)F_k(B_j)\tilde\Omega_k\right)
=\lim_{L\to\infty} \omega\bigl( F_{L,k}(A_i)F_{L,k}(B_j) \bigr)
\end{equation}
i.e. essentially determined by the two-point functions of the given state $\omega$.
Therefore as is clear from (\ref{e:scal-prod}), all $(n+m)$-point functions are given by
the two-point function (\ref{e:two-pt}). The definition (\ref{e:scal-prod}) defines
completely the fluctuation operators $F_k(A)$ on the Hilbert space
$\tilde{\mathcal{H}}_k$.

On the other hand, the non-commutative law of large numbers, here of large operators,
leads straightforwardly to the canonical commutation relation
\begin{equation*}
\lim_{L\to\infty} \bigl[ F_{L,k}(n),F_{L,k}(o) \bigr] 
=\lim_{L\to\infty}\frac{1}{2V}\int_\Lambda dx\;[n(x),o(x)]= \omega([n(x),o(x)])
=i\sqrt{\rho_0},
\end{equation*}
or
\begin{equation}\label{eq:fluc-com}
\bigl[ F_k(n),F_k(o) \bigr] = i\sqrt{\rho_0}.
\end{equation}
This is the basic result for the definition of the phase operator. Equation
(\ref{eq:fluc-com}) means that the number fluctuation operator $F_k(n)$ and the order
parameter fluctuation operator $F_k(o)$ are canonically conjugate (compare with
$[q,p]=i\hbar$). Hence for the physics of BEC we found on the level of fluctuations the
canonical pair $\bigl(F_k(n),F_k(o)\bigr)$. Clearly the operator $F_k(o)$ satisfies all
basic physical requirements for playing the role of what is usually called the {\em phase
operator} of the condensate.

The reader will have recognised from (\ref{e:scal-prod}) and (\ref{eq:fluc-com}) that the
fluctuation operators 
\begin{equation*}
F_k(A),F_k(B),\dotsc
\end{equation*}
form an algebra of Boson field operators 
and in particular that $F_k(n)$ and $F_k(o)$ form quantum canonical variables with a
quantisation parameter $\sqrt{\rho_0}$ (compare with $\hbar$). On the other hand, from  
(\ref{e:scal-prod}) it is clear that the vector $\tilde\Omega_k$ defines a generalised or
quasi free (gaussian) state
\begin{equation*}
\tilde\omega_k(\cdot)=\left(\tilde\Omega_k, \cdot\; \tilde\Omega_k\right)
\end{equation*}
on the Boson field algebra (see \cite{bratteli:1996}).

This means that the central limit theorem for the $k$-mode fluctuations in the state
$\omega$ defines an equilibrium state $\tilde\omega_k$ on the fluctuation operators. The
mentioned quasi free character means that all correlation functions of limit fluctuation
operators are polynomial functions only of the one- and two-point functions. For more
details we refer once more to \cite{michoel:1999b}. 

The reader might ask for the unicity of this {\em phase operator}, if being defined only
as the canonically adjoint operator to the number fluctuation operator. The interested
reader is referred to section \ref{s:rel-ops} and 
\cite{michoel:1999b, michoel:2000} for that discussion.

\section{The model and equilibrium states}\label{s:model}

We consider two coupled Bose-Einstein condensates, each of them modelled by an imperfect
or mean field Bose gas. 
Denote $a^\sharp_i(x)$, $i=1,2$ the creation and annihilation operators for the two
Bose gases, i.e.
\begin{equation*}
[a_i(x), a^*_j(y)]=\delta_{i,j}\delta(x-y).
\end{equation*}
We assume that the two gases have the same particle density $\frac{\rho}{2}$ (hence
$\rho$ is the total particle density $\rho=\frac{1}{V}\omega_L(N_L)$ with 
$N_L=N_{1,L}+N_{2,L}$), and also that they are of the same type of particles (i.e. there
is only one mean field constant $\lambda$).
We also assume a phase difference $\varphi$ between the gases, and 
model the Josephson coupling between the gases by a term
\begin{equation}\label{eq:joseph-coupl}
C^{1,2}_L = -\gamma \sum_{k\in\Lambda^*} a^*_{1,k}a_{2,k}e^{-i\varphi} + a^*_{2,k}a_{1,k}
e^{i\varphi},
\end{equation}
with $\gamma >0$ the coupling constant.

Hence the local Hamiltonian of the system we study is given by:
\begin{align}\label{eq:ham}
H_L &= T_{1,L} + T_{2,L} - \mu_L N_L + \frac{\lambda}{2 V} N_L^2 + C^{1,2}_L \nonumber\\
&=\sum_{k \in \Lambda^*} (\epsilon_k - \mu_L) (a^*_{1,k}a_{1,k} + a^*_{2,k}a_{2,k} )
 + \frac{\lambda}{2 V}(N_{1,L}+N_{2,L})^2 \nonumber\\
&\quad-\gamma \sum_{k\in\Lambda^*} a^*_{1,k}a_{2,k}e^{-i\varphi} + a^*_{2,k}a_{1,k}
e^{i\varphi}.
\end{align}

In this section we find the limiting Gibbs states $\omega =
\lim_{L\to\infty}\omega_L$ at inverse temperature $\beta$ of this model. 
A rigorous study along the lines of
\cite{fannes:1980b} is perfectly possible, but we permit ourselves here a more intuitive
approach. As in any mean field model, we replace the Hamiltonian (\ref{eq:ham}) by a
state dependent effective Hamiltonian
\begin{equation}\label{eq:ham-eff}
H_L^\omega = \sum_{k \in \Lambda^*} (\epsilon_k -\mu +\lambda \rho)
(a^*_{1,k}a_{1,k} + a^*_{2,k}a_{2,k})
-\gamma \sum_{k\in\Lambda^*} a^*_{1,k}a_{2,k}e^{-i\varphi} + a^*_{2,k}a_{1,k}e^{i\varphi}
\end{equation}
with $\mu=\lim_{L\to\infty}\mu_L$ in correspondence with the constraint 
$\rho = \frac{1}{V}\omega(N_L)$. This effective 
Hamiltonian is bilinear in the
creation and annihilation operators and therefore it can be diagonalized by a Bogoliubov
transformation.

Let $\delta^\omega_L(\cdot) = [H_L^\omega,\cdot]$ be the generator of this dynamics
and $f_k = \epsilon_k -\mu +\lambda \rho$. Then
\begin{equation*}
\delta^\omega_L \begin{pmatrix} a^*_{1,k} \\ a^*_{2,k} \end{pmatrix} = 
\begin{pmatrix} f_k & -\gamma e^{i\varphi}\\
               -\gamma e^{-i\varphi} & f_k \end{pmatrix}
\begin{pmatrix} a^*_{1,k} \\ a^*_{2,k} \end{pmatrix}.
\end{equation*}
The matrix
\begin{equation*}
\begin{pmatrix} f_k & -\gamma e^{i\varphi}\\
               -\gamma e^{-i\varphi} & f_k \end{pmatrix}
\end{equation*}
has eigenvalues $E_k^\pm$, 
\begin{equation*}
E_k^\pm = f_k \pm \gamma,
\end{equation*}
with corresponding eigenoperators $b^*_{\pm,k}$,
\begin{equation}\label{eq:b-ops}
b^*_{\pm,k} = \frac{1}{\sqrt{2}}\bigl( a^*_{1,k}e^{-\frac{i}{2}\varphi} \mp a^*_{2,k} 
e^{\frac{i}{2}\varphi} \bigr),
\end{equation}
i.e.
\begin{equation}\label{eq:time-evol}
\delta^\omega_L(b^*_{\pm,k}) = E_k^\pm b^*_{\pm,k}.
\end{equation}
The $b^\sharp_\pm(x)$ still satisfy Boson commutation rules. The energy spectrum of the
quasi-particles $b^\sharp_\pm(x)$ has two branches, $\{E_k^\pm|k\in\R^\nu\}$.

By a standard argument, using the Boson commutation rules and the correlation inequalities
\cite{fannes:1977, fannes:1977b}, characterizing the limit equilibrium states,
\begin{equation}\label{eq:cor-ineq}
\lim_{L\to\infty} \beta \omega\bigl( X^* \delta^\omega_L(X) \bigr) \geq \omega(X^*X)
\ln\frac{\omega(X^*X)}{\omega(XX^*)},
\end{equation}
for $X$ any polynomial in the creation and annihilation operators, one finds in a
straightforward manner:
\begin{equation}\label{eq:2-pt-cor}
\omega(b^*_{\pm,k}b_{\pm,k}) = \frac{1}{e^{\beta E^\pm_k}-1}.
\end{equation}

Along the usual lines of the derivation of Bose-Einstein condensation, we find a
critical density (or inverse temperature) above which one derives the following value 
of the chemical potential:
\begin{equation*}
\mu = \lambda \rho - \gamma,
\end{equation*}
i.e. $E_k^- = \epsilon_k$ and $E_k^+>2\gamma$ for all $k$. Hence there is a macroscopic 
occupation of the 
0-momentum state of the `$-$'-mode possible; the condensate density is given by:
\begin{equation}\label{eq:-mode-condens}
\lim_{V\to\infty}\frac{1}{V}\omega(b^*_{-,0}b_{-,0}) = \rho_0 > 0.
\end{equation}
There is no condensation for the `+'-mode since $\lim_{k\to 0}E_k^+ = 2\gamma >0$ for
$\mu = \lambda \rho - \gamma$. 

From $\rho = \lim_{L\to\infty}V^{-1}\omega(N_L)$ and (\ref{eq:b-ops}), one finds
\begin{align*}
\rho &= \lim_{L\to\infty}\frac{1}{V}\sum_k \omega\bigl(b^*_{-,k}b_{-,k}\bigr)+
\omega\bigl(b^*_{+,k}b_{+,k}\bigr)\\
&=\rho_0 + \int_{\R^\nu}\frac{dk}{(2\pi)^3}\frac{1}{e^{\beta\epsilon_k}-1}
+ \int_{\R^\nu}\frac{dk}{(2\pi)^3}\frac{1}{e^{\beta(\epsilon_k+2\gamma)}-1}.
\end{align*}
For $\alpha\geq 0$, denote
\begin{equation}\label{eq:rho-alpha}
\rho(\alpha)=\int_{\R^\nu}\frac{dk}{(2\pi)^3}\frac{1}{e^{\beta(\epsilon_k+\alpha)}-1},
\end{equation}
then (\ref{eq:-mode-condens}) becomes
\begin{equation*}
\rho_0 = \rho - \rho(0) - \rho(2\gamma).
\end{equation*}
It is clear that $\rho(0)+\rho(2\gamma)$ is the critical density.

The Bose-Einstein condensation (\ref{eq:-mode-condens}) implies a spontaneous breaking of
the gauge symmetry, i.e. the limiting Gibbs state $\omega$ decomposes with respect to the
$U(1)$ gauge group into distinct extremal equilibrium states:
\begin{equation*}
\omega = \frac{1}{2\pi}\int_0^{2\pi}d\theta\; \omega_\theta,
\end{equation*}
where each of the states $\omega_\theta$ is determined by the two-point function 
(\ref{eq:2-pt-cor}) and the one-point function
\begin{equation}\label{eq:1-pt-cor}
\omega_\theta \bigl( b^*_-(x) \bigr) = \sqrt{\rho_0}e^{i\theta}.
\end{equation}
This important point, whose non-triviality is often overlooked, is proved in 
\cite{fannes:1982}. Of course, the other way round, namely that 
gauge symmetry breaking implies
condensation, is well known and follows trivially from the Schwartz inequality.

From now on we choose a particular extremal equilibrium state $\omega_\theta$, and
without loss of generality we take $\theta=0$. For notational simplicity,
this state is again denoted by $\omega$.
Using once more the correlation inequalities (\ref{eq:cor-ineq}), it is not difficult to
show that the higher order correlations decompose into sums and products of one- and 
two-point 
correlations, given by (\ref{eq:1-pt-cor}) respectively (\ref{eq:2-pt-cor}), i.e. the
state $\omega$ is quasi-free. Therefore we have completely characterized the equilibrium
states. 

It is clear that the gauge symmetry breaking state under discussion indeed corresponds to 
a state of two different condensates interacting through a Josephson coupling
(\ref{eq:joseph-coupl}). Since $\omega(b^*_+(x))=0$ and $\omega(b^*_-(x)) =
\sqrt{\rho_0}$, we find using (\ref{eq:b-ops})
\begin{equation}\label{eq:phase-diff}
\omega(a_1^*(x))e^{-\frac{i}{2}\varphi}=\omega(a_2^*(x))e^{\frac{i}{2}\varphi}=\sqrt{
\frac{\rho_0}{2}}.
\end{equation}
Notice that $\varphi$ is indeed the phase difference between the condensates. Our 
arbitrary choice $\theta=0$ then actually determines both phases to be 
$\pm\frac{\varphi}{2}$ and the choice of equal
particle densities $\frac{\rho}{2}$ yields equal condensate densities $\frac{\rho_0}{2}$. 
Moreover it is obvious that the gapless mode $E_k^-$ is related to the broken gauge
symmetry, i.e. to the Bose-Einstein condensation, and that the mode $E_k^+$ with energy 
gap $2\gamma$ arises due to the presence of the Josephson coupling.
The detailed study of the fluctuation operators corresponding to these two 
excitation branches and modes is the subject of the subsequent sections.

\section{Total number and phase fluctuation operators}\label{s:total-ops}
Motivated by the discussion in section \ref{s:fluc-op} and \cite{michoel:1999b},
define the total number and phase fluctuations in the box $\Lambda$ by ($k\not= 0$)
\begin{align}
F_{L,k}(n_{tot}) &= \frac{1}{\sqrt{V}}\int_\Lambda dx \bigl( a_1^*(x)a_1(x)+a_2^*(x)a_2(x)
  -\rho\bigr) \cos k.x \label{eq:tot-num-fluc}\\
F_{L,k}(\phi_{tot}) &= \frac{i}{\sqrt{V}}\int_\Lambda dx \bigl( b_-^*(x)-b_-(x)\bigr)
\cos k.x,\label{eq:tot-phase-fluc}
\end{align}
where $b_-(x)$ is defined in (\ref{eq:b-ops}).

Again on the basis of the law of large numbers:
\begin{equation}\label{eq:tot-fluc-comm}
\lim_{L\to\infty} \bigl[ F_{L,k}(n_{tot}),F_{L,k}(\phi_{tot}) \bigr] = \lim_{L\to\infty}
\frac{i}{2\sqrt{V}}(b^*_{-,0}+b_{-,0})=i\sqrt{\rho_0}.
\end{equation}

Using (\ref{eq:b-ops}) one writes also
\begin{equation*}
a_1^*(x)a_1(x)+a_2^*(x)a_2(x) = b^*_-(x)b_-(x) + b^*_+(x)b_+(x),
\end{equation*}
or
\begin{equation*}
F_{L,k}(n_{tot}) = F_{L,k}(n_-) + F_{L,k}(n_+),
\end{equation*}
with $n_\pm$ defined in the obvious sense. Hence the total number operator fluctuation is
the sum of two number operator fluctuations of two imperfect Bose gases.
For a single imperfect Bose gas (see section \ref{s:fluc-op}), one finds this 
analysis as the subject of 
\cite{michoel:1999b}, making the present analysis straightforward.

As already discussed in section \ref{s:fluc-op}, and apparent from (\ref{eq:tot-fluc-comm}), the
limiting fluctuation operators $F_k(\cdot)$ satisfy Bosonic commutation rules (although
the \emph{local} fluctuation operators \emph{do not}). 
Equation (\ref{eq:tot-fluc-comm}) learns also that the fluctuation operators 
$F_k(n_{tot})$ and $F_k(\phi_{tot})$ constitute a canonical pair, generating an algebra of
canonical commutation relations (\emph{CCR}) of fluctuation observables of the system.
Furthermore, it is shown in \cite{goderis:1989b, goderis:1990} and briefly discussed in
section \ref{s:fluc-op} that the central limit theorem also fixes an
equilibrium state $\tilde\omega_k$ on this algebra of limiting fluctuation operators, 
which is a \emph{CCR}-algebra of Bosonic field operators. This state is shown
to be quasi-free and gauge invariant, and hence  completely 
determined by its two-point function, given by \cite{goderis:1989b, goderis:1990}:
\begin{equation*}
\tilde\omega_k\bigl( F_k(A)F_k(B) \bigr) = \lim_{L\to\infty}\int_\Lambda dz\;
\omega\bigl( A(z) B(0) \bigr) \cos k.z,
\end{equation*}
where $A,B$ are (in the present case) polynomials in the microscopic canonical 
Bosonic field  
operators. Remark that there are no technical problems related to the central limit
theorem for $k\not=0$, off-diagonal long-range order correlations do appear only at $k=0$.

The first step in our study of the total number- and phase fluctuation operators is 
to determine their variances.
\begin{proposition}
For $k\not=0$ we have
\begin{align}
\begin{split}
(i) \quad \tilde\omega_k\bigl(F_k(n_{tot})^2\bigr) &= \tilde\omega_k\bigl(F_k(n_-)^2
\bigr) +\tilde\omega_k\bigl(F_k(n_+)^2\bigr)\\
 &= \frac{\rho_0}{2}\coth\frac{\beta\epsilon_k}{2}+\frac{1}{2}\int_{\R^\nu}\frac{dp}{(2\pi)^3}
\frac{e^{\beta\epsilon_{p+k}}+e^{\beta\epsilon_p}}{\bigl(e^{\beta\epsilon_{p+k}}-1\bigr)
\bigl(e^{\beta\epsilon_p}-1\bigr)}\\
&+\frac{1}{2}\int_{\R^\nu}\frac{dp}{(2\pi)^3}
\frac{e^{\beta E^+_{p+k}}+e^{\beta E^+_p}}{\bigl(e^{\beta E^+_{p+k}}-1\bigr)
\bigl(e^{\beta E^+_p}-1\bigr)}
\end{split}\label{eq:var-tot-num-fluc}\\
(ii) \quad \tilde\omega_k\bigl(F_k(\phi_{tot})^2\bigr) &= \frac{1}{2}
\coth\frac{\beta\epsilon_k}{2}.\label{eq:var-tot-phase-fluc}
\end{align}
\end{proposition}
\begin{proof}
$(i)$ The fact that $\tilde\omega_k\bigl(F_k(n_{tot})^2\bigr) = 
\tilde\omega_k\bigl(F_k(n_-)^2\bigr) +\tilde\omega_k\bigl(F_k(n_+)^2\bigr)$ follows 
from the fact that the `$+$'- and 
`$-$'-mode are independent of each other. For a calculation of the explicit expression for
$\tilde\omega_k\bigl(F_k(n_{tot})^2\bigr)$, see \cite{michoel:1999b}.\\
$(ii)$ This is simply the two-point function of the state $\omega$, 
see \cite{michoel:1999b}.
\end{proof}
Remark that the result of this proposition, supplemented with (\ref{eq:tot-fluc-comm}) is 
sufficient in order to characterize completely the limiting fluctuation operators 
$F_k(n_{tot})$ and $F_k(\phi_{tot})$ on a well defined Hilbert space (see section
\ref{s:fluc-op}). We do not enter into these technical details.

Of course we are interested particularly in the limit $k\to 0$ of these operators, in
order to see how the long-range order due to the Bose-Einstein condensation manifests 
itself on the level of the fluctuations. In \cite{michoel:1999b} one can find a discussion
demonstrating that quantum effects will only be present in the limit $k\to 0$ if one
works in the groundstate $\omega^g$, defined as the zero-temperature limit of the
equilibrium state $\omega$:
\begin{equation*}
\omega^g = \lim_{\beta\to\infty}\omega.
\end{equation*}
At $T=0$, the variances (\ref{eq:var-tot-num-fluc}) and (\ref{eq:var-tot-phase-fluc})
simplify to
\begin{align*}
\tilde\omega^g_k\bigl(F_k(n_{tot})^2\bigr) &= \frac{\rho_0}{2}\\
\tilde\omega^g_k\bigl(F_k(\phi_{tot})^2\bigr) &= \frac{1}{2},
\end{align*}
and the limit $k\to 0$ is trivial:
\begin{equation*}
F_0(n_{tot})=\lim_{k\to 0}F_k(n_{tot}) \quad\text{and}\quad
F_0(\phi_{tot})=\lim_{k\to 0}F_k(\phi_{tot}).
\end{equation*}
These are well defined fluctuation operators, satisfying
\begin{align*}
\bigl[ F_0(n_{tot}),F_0(\phi_{tot}) \bigr] &= i\sqrt{\rho_0}\\
\tilde\omega^g_0\bigl(F_0(n_{tot})^2\bigr) &= \frac{\rho_0}{2}\\
\tilde\omega^g_0\bigl(F_0(\phi_{tot})^2\bigr) &= \frac{1}{2}.
\end{align*}

We derived the exact uncertainty relation between the number operator and phase operator,
given by
\begin{equation*}
\tilde\omega^g_0\bigl(F_0(n_{tot})^2\bigr)\tilde\omega^g_0\bigl(F_0(\phi_{tot})^2\bigr)
=\frac{\rho_0}{4}.
\end{equation*}
Remark that the condensate density $\rho_0$, in fact $\sqrt{\rho_0}$, acts in this equation,
as well as in equation (\ref{eq:tot-fluc-comm}), as a quantisation parameter (compare with
$\hbar$). Consequently, the whole content of our results, as well as all physical
interpretations, do disappear in the absence of condensation, i.e. if $\rho_0=0$.

Finally remark that we have omitted in our notation the state 
dependence of the fluctuation operators 
throughout, although this dependence is important. Fluctuation operators corresponding to
different states (e.g. corresponding to different temperature or different phase) in fact 
are not comparable as they act on completely different Hilbert spaces. We do not enter
into these mathematical subtleties.

\section{Relative number and phase fluctuation operators}\label{s:rel-ops}

The relative number operator in a finite volume is:
\begin{equation*}
N_{L,rel}=N_{1,L} - N_{2,L},
\end{equation*}
and its $k$-mode fluctuation
\begin{equation*}
F_{L,k}(n_{rel}) = \frac{1}{V^{1/2}}\int_{\Lambda}dx\; \bigl( a^*_1(x)a_1(x) - 
a^*_2(x)a_2(x) \bigr) \cos k.x.
\end{equation*}
As before, we are primarily interested in the limit $L\to\infty$, followed
by the limit $k\to 0$. 
The relative number operator $N_{L,rel}$ is not the generator of a symmetry of the local
Hamiltonian $H_L$ (\ref{eq:ham}) because of the Josephson coupling term. This means that
there is no question of spontaneous symmetry breaking for the relative number operator, as
it is not a symmetry. Furthermore a straightforward computation of the dynamics of 
$N_{L,rel}$ learns that its spectrum belongs to the excitation branch $E^+ - E^-$. 
Since $E_k^+ - E_k^-=2\gamma >0$, the spectrum of $N_{L,rel}$ shows an
energy gap. Hence
the $0$-mode fluctuations of the relative number operator and its adjoint will be normal.
Therefore the limits $L\to\infty$ and $k\to 0$ may be interchanged and hence the starting
point of our investigation can be the operator (i.e. the case $k=0$):
\begin{equation}\label{eq:rel-num-fluc}
F_L(n_{rel}) = \frac{1}{V^{1/2}}\int_{\Lambda}dx\; \bigl( a^*_1(x)a_1(x) - 
a^*_2(x)a_2(x) \bigr) = \frac{1}{V^{1/2}}\sum_k a^*_{1,k}a_{1,k} - a^*_{2,k}a_{2,k}.
\end{equation}
Next define the fluctuation operator of the relative current:
\begin{align}\label{eq:rel-cur-fluc}
F_L(j_{rel}) &= \frac{i}{2\gamma V^{1/2}}\int_{\Lambda}dx\;\bigl(a^*_1(x)a_2(x)e^{-i\varphi}-
a^*_2(x)a_1(x)e^{i\varphi} \bigr) \nonumber\\
&=\frac{i}{2\gamma V^{1/2}}\sum_k a^*_{1,k}a_{2,k} e^{-i\varphi} - a^*_{2,k}a_{1,k}
e^{i\varphi}.
\end{align}
This operator clearly corresponds to the 0-mode fluctuations of the relative current from 
one gas into the other. Below, we show in a series of steps 
that the operators (\ref{eq:rel-num-fluc}) and
(\ref{eq:rel-cur-fluc}) are each others adjoint in the limit $L\to\infty$. 
Afterwards we show the relations between the relative current fluctuation 
operator on the one hand and the relative phase fluctuation operator on the other hand.

A central limit theorem and reconstruction theorem can again be proved for these
operators (see \cite{goderis:1990, michoel:1999b}), proving the existence of the Bosonic
field operators
\begin{align*}
F(n_{rel}) &= \lim_{L\to\infty}F_L(n_{rel})\\
F(j_{rel}) &= \lim_{L\to\infty}F_L(j_{rel})
\end{align*}
in a rigorous mathematical sense.

Let $\delta_\omega(\cdot)=\lim_{L\to\infty}[H_L^\omega,\cdot]$ be the infinitesimal 
generator of the dynamics ($H_L^\omega$ is the effective Hamiltonian defined in 
equation (\ref{eq:ham-eff}), with $\mu = \lambda \rho - \gamma$). 

In order to prove the
properties below, it is convenient to write (\ref{eq:rel-num-fluc}) and 
(\ref{eq:rel-cur-fluc}) in terms of the quasi-particle operators (\ref{eq:b-ops}):
\begin{align}
F_L(n_{rel}) &= \frac{1}{V^{1/2}}\sum_k b^*_{+,k}b_{-,k} + b^*_{-,k}b_{+,k} \\
F_L(j_{rel}) &= \frac{i}{2\gamma V^{1/2}}\sum_k b^*_{+,k}b_{-,k} -
b^*_{-,k}b_{+,k}\label{eq:rel-cur-fluc-2}.
\end{align}

\begin{proposition}\label{pr:rel-fluc-can-pair}
The operators $F(n_{rel})$ and $F(j_{rel})$ form a canonical pair and satisfy
\begin{equation*}
\bigl[ F(n_{rel}),F(j_{rel}) \bigr] = i c_{rel},
\end{equation*}
where
\begin{equation*}
c_{rel} = \beta \lim_{L\to\infty} \bigl( F_L(n_{rel}),F_L(n_{rel}) \bigr)_\sim 
= \frac{1}{\gamma}\bigl(\rho_0 + \rho(0) - \rho(2\gamma)\bigr) > 0;
\end{equation*}
$\rho(\alpha)$ is defined in (\ref{eq:rho-alpha}), and $(\cdot,\cdot)_\sim$ is the Duhamel
two-point function defined below.
\end{proposition}
\begin{proof}
The first statement follows again from the general theory on normal fluctuation operators
\cite{goderis:1990,michoel:1999b}. Also, it is an easy calculation to show that
\begin{equation*}
\bigl[ H_L^\omega, F_L(j_{rel})\bigr]=iF_L(n_{rel}).
\end{equation*}
Because of the presence of the energy gap $2\gamma>0$, this can be written as:
\begin{equation*}
F_L(j_{rel}) = i \delta_\omega^{-1}\bigl( F_L(n_{rel})\bigr).
\end{equation*}
Therefore, using the fact that $\omega$ is an equilibrium (\emph{KMS}) state, one gets
\begin{align*}
\bigl[ F(n_{rel}),F(j_{rel}) \bigr] &= \lim_{L\to\infty} \omega\bigl( 
\bigl[ F_L(n_{rel}),F_L(j_{rel}) \bigr]\bigr)\\
&=i \lim_{L\to\infty} \omega\bigl(F_L(n_{rel}) [1-e^{-\beta \delta_\omega}]
\delta_\omega^{-1} F_L(n_{rel})\bigr)\\
&=i\beta \lim_{L\to\infty} \bigl( F_L(n_{rel}),F_L(n_{rel}) \bigr)_\sim.
\end{align*}
The explicit expression for the Duhamel two-point function 
$\bigl( F_L(n_{rel}),F_L(n_{rel}) \bigr)_\sim$ then follows from
\begin{equation*}
\omega\bigl( \bigl[ F_L(n_{rel}),F_L(j_{rel}) \bigr]\bigr) 
= \frac{i}{\gamma V}\sum_k\omega\bigl( b^*_{-,k}b_{-,k}-b^*_{+,k}b_{+,k}\bigr)
\xrightarrow{L\to\infty}\frac{i}{\gamma}\bigl(\rho_0 + \rho(0) - \rho(2\gamma)\bigr).
\end{equation*}
\end{proof}
 
The infinitesimal generator $\delta_\omega$ of the microdynamics induces a natural
infinitesimal generator $\tilde\delta_\omega$ of a dynamics on the macroscopic fluctuation
operators by the formula \cite{goderis:1990}:
\begin{equation*}
\tilde\delta_\omega F(A)=F(\delta_\omega(A)).
\end{equation*}
\begin{proposition}\label{pr:dyn}
The infinitesimal generator $\tilde\delta_\omega$ on the
macroscopic fluctuations is given by:
\begin{align}
\tilde\delta_\omega F(n_{rel}) &= -i (2\gamma)^2 F(j_{rel})\label{eq:motion-rel-num}\\
\tilde\delta_\omega F(j_{rel}) &= i F(n_{rel})\label{eq:motion-rel-cur}.
\end{align}
Hence $F(n_{rel})$ and $F(j_{rel})$ are eigenvectors of $\tilde\delta_\omega^2$:
\begin{equation*}
\tilde\delta_\omega^2 F(n_{rel}) = (2\gamma)^2F(n_{rel}) \;,\;
\tilde\delta_\omega^2 F(j_{rel}) = (2\gamma)^2F(j_{rel}),
\end{equation*}
yielding the macrodynamics $\tilde\alpha_t$ on the fluctuation operators:
\begin{align*}
\tilde\alpha_t F(n_{rel}) &= e^{it \tilde\delta_\omega}F(n_{rel}) = F(n_{rel})
\cos(2\gamma t) + (2\gamma)F(j_{rel})\sin(2\gamma t)\\
\tilde\alpha_t F(j_{rel}) &= e^{it \tilde\delta_\omega}F(j_{rel}) = 
\frac{1}{2\gamma}F(n_{rel})\sin(2\gamma t) + F(j_{rel})\cos(2\gamma t).
\end{align*}
\end{proposition}
\begin{proof}
This follows immediately from the relations
\begin{align*}
\bigl[ H_L^\omega, F_L(j_{rel})\bigr]&=iF_L(n_{rel})\\
\bigl[ H_L^\omega, F_L(n_{rel})\bigr]&=-i(2\gamma)^2 F_L(j_{rel}). 
\end{align*}
\end{proof}

Remark that we proved that the pair of variables $\bigl( F(n_{rel}),F(j_{rel}) \bigr)$ is
dynamically independent from the other variables of the system. The pair behaves
dynamically as a pair of quantum oscillator variables with a frequency equal to $2\gamma$.

\begin{proposition}[Virial theorem]\label{pr:vir-thm}
The mean square fluctuation of the relative number operator is proportional to the 
mean square fluctuation of the relative current operator, in particular:
\begin{equation*}
\tilde \omega\bigl( F(n_{rel})^2 \bigr) = (2\gamma)^2 \tilde\omega\bigl( F(j_{rel})^2 \bigr).
\end{equation*}
\end{proposition}
\begin{proof}
This follows from the time invariance of $\tilde\omega$, i.e. $\tilde\omega \circ
\tilde\delta_\omega = 0$:
\begin{equation*}
0=\tilde\omega\bigl(\tilde\delta_\omega[F(n_{rel})F(j_{rel})] \bigr) =
\tilde\omega\bigl(\tilde\delta_\omega[F(n_{rel})]F(j_{rel}) \bigr) + 
\tilde\omega\bigl(F(n_{rel})\tilde\delta_\omega[F(j_{rel})]\bigr),
\end{equation*}
and the equations of motion (\ref{eq:motion-rel-num}) and (\ref{eq:motion-rel-cur}).
\end{proof}

\begin{proposition}\label{pr:var-rel}
The mean square fluctuation of the relative number operator is given by
\begin{equation*}
\tilde \omega\bigl( F(n_{rel})^2 \bigr) = c_{rel}\gamma \coth \beta\gamma.
\end{equation*}
\end{proposition}
\begin{proof}
We compute this quantity using the correlation inequalities (\ref{eq:cor-ineq}), rewritten
in the form
\begin{equation}\label{eq:cor-ineq-2}
\frac{-\beta\omega\bigl(X\delta_\omega(X^*)\bigr)}{\omega(XX^*)} \leq
\ln\frac{\omega(X^*X)}{\omega(XX^*)} \leq 
\frac{\beta\omega\bigl(X^*\delta_\omega(X)\bigr)}{\omega(X^*X)}.
\end{equation}
We take for $X$ the operator $A_L=F_L(n_{rel})+i(2\gamma)F_L(j_{rel})$ and then let 
$L\to\infty$, and use proposition \ref{pr:rel-fluc-can-pair}.

One gets
\begin{equation*}
\lim_{L\to\infty}\omega(A_L A_L^*) = \tilde\omega\bigl(F(n_{rel})^2\bigr) + 
(2\gamma)^2 \tilde\omega\bigl(F(j_{rel})^2\bigr) + (2\gamma)c_{rel},
\end{equation*}
and by the virial theorem (proposition \ref{pr:vir-thm}):
\begin{equation*}
\lim_{L\to\infty}\omega(A_L A_L^*) = 2 \tilde\omega\bigl(F(n_{rel})^2\bigr) + 
(2\gamma)c_{rel}.
\end{equation*}
Analogously
\begin{equation*}
\lim_{L\to\infty}\omega(A_L^* A_L) = 2 \tilde\omega\bigl(F(n_{rel})^2\bigr) - 
(2\gamma)c_{rel}.
\end{equation*}
On the other hand,
\begin{equation*}
\delta_\omega(A_L)=-i(2\gamma)^2 F_L(j_{rel}) -(2\gamma)F_L(n_{rel})=-(2\gamma) A_L,
\end{equation*}
and hence
\begin{align*}
\lim_{L\to\infty}\omega\bigl(A_L^* \delta_\omega(A_L)\bigr)&=-4\gamma
\tilde\omega\bigl(F(n_{rel})^2\bigr)+(2\gamma)^2c_{rel}\\
\lim_{L\to\infty}\omega\bigl(A_L \delta_\omega(A^*_L)\bigr)&=4\gamma
\tilde\omega\bigl(F(n_{rel})^2\bigr)+(2\gamma)^2c_{rel}.
\end{align*}
After substitution in (\ref{eq:cor-ineq-2}) one gets
\begin{equation*}
\ln\frac{2 \tilde\omega\bigl(F(n_{rel})^2\bigr) - (2\gamma)c_{rel}}
{2 \tilde\omega\bigl(F(n_{rel})^2\bigr) + (2\gamma)c_{rel}}=-2\beta\gamma,
\end{equation*}
or alternatively
\begin{equation*}
\tilde \omega\bigl( F(n_{rel})^2 \bigr) = c_{rel}\gamma \coth \beta\gamma.
\end{equation*}
\end{proof}

This finishes the complete study of the static and dynamic properties of the canonical
pair $\bigl(F(n_{rel}),F(j_{rel})\bigr)$ of the relative density and current fluctuations.
We proved rigorously that for all temperatures below the condensation temperature and with
non-zero condensate, this pair behaves like a pair of quantum harmonic oscillator
variables, describing oscillations of the fluid from type 1 into type 2 and vice versa,
yielding the typical interference pattern.
The plasmon frequency is given by $2\gamma$. All this is physically clear.

Our next and final problem is to find out what the position of the phase is in all this.
We turn our attention now to look for a relation between the relative current fluctuation
operator $F(j_{rel})$ and the relative phase fluctuation operator, which we define in an
analogous form as the total phase fluctuation operator, as follows:
\begin{equation*}
F_L(\phi_{rel})=\frac{i}{2V^{1/2}}\int_\Lambda dx\; \bigl( b^*_+(x) - b_+(x) \bigr) =
\frac{i}{2}\bigl(b^*_{+,0} - b_{+,0}\bigr).
\end{equation*}
Denote its central limit by $F(\phi_{rel})$.

First, observe that one can distinguish two terms in the relative current fluctuation 
(\ref{eq:rel-cur-fluc-2}), namely the $k=0$ part and the rest:
\begin{equation*}
F_L(j_{rel}) = \frac{i}{2\gamma V^{1/2}} \bigl(b^*_{+,0}b_{-,0} - b^*_{-,0}b_{+,0}\bigr)
+\frac{i}{2\gamma V^{1/2}}\sum_{k\not=0}b^*_{+,k}b_{-,k} - b^*_{-,k}b_{+,k}.
\end{equation*}
Denote the first term by
\begin{equation*}
F_L(j^0_{rel})=\frac{i}{2\gamma V^{1/2}} \bigl(b^*_{+,0}b_{-,0} - b^*_{-,0}b_{+,0}\bigr),
\end{equation*}
and its central limit by $F(j^0_{rel})$. Also denote by $\omega^g$ the ground state, 
obtained as the zero-temperature limit of the equilibrium state $\omega$:
\begin{equation*}
\omega^g(\cdot)= \lim_{\beta\to\infty}\omega(\cdot),
\end{equation*}
and $\tilde\omega^g$ the corresponding ground state for the limiting fluctuation operators
observables:
\begin{equation*}
\tilde\omega^g(\cdot)= \lim_{\beta\to\infty}\tilde\omega(\cdot).
\end{equation*}
\begin{proposition}
We have the following relationships between the limiting fluctuation operators:
\begin{enumerate}
\item $\forall \beta>0$, $\beta=\infty$ included, 
\begin{equation*}
F(j^0_{rel}) = \frac{\sqrt{\rho_0}}{\gamma} F(\phi_{rel});
\end{equation*} 

\item for $\beta=\infty$,
\begin{equation*}
F(j_{rel}) = F(j^0_{rel}) = \frac{\sqrt{\rho_0}}{\gamma} F(\phi_{rel}).
\end{equation*}
\end{enumerate}
\end{proposition}
\begin{proof}
As shown in \cite{goderis:1989b,goderis:1990}, two fluctuation operators $F(A),F(B)$ are 
equal in the algebra of fluctuation operators whenever
\begin{equation}\label{eq:coarse-grain}
\tilde\omega\bigl(F(A-B)^2\bigr) = 0,
\end{equation}
i.e. whenever the variance of the difference $A-B$ of the operators vanishes. This is
expressing in a mathematical rigorous setting, the phenomenon of coarse graining on the
level of fluctuations.

Therefore we calculate
\begin{equation*}\begin{split}
\tilde\omega\Bigl(\bigl[F(j^0_{rel})-\frac{\sqrt{\rho_0}}{\gamma}
F(\phi_{rel})\bigr]^2 \Bigr) 
&= \tilde\omega\bigl( F(j^0_{rel})^2 \bigr) + \frac{\rho_0}{\gamma^2} 
\tilde\omega\bigl(F(\phi_{rel})^2\bigr)\\ &- \frac{\sqrt{\rho_0}}{\gamma}
\tilde\omega\bigl( F(j^0_{rel})F(\phi_{rel})\bigr)- \frac{\sqrt{\rho_0}}{\gamma}
\tilde\omega\bigl( F(\phi_{rel})F(j^0_{rel})\bigr).
\end{split}
\end{equation*}
One finds, using the explicit knowledge of the state $\omega$ (section \ref{s:model}):
\begin{align*}
\tilde\omega\bigl( F(j^0_{rel})^2 \bigr) &=\tilde\omega\bigl(F(\phi_{rel})^2\bigr)=
 \frac{\rho_0}{4\gamma^2}\coth\beta\gamma \\
\tilde\omega\bigl( F(j^0_{rel})F(\phi_{rel})\bigr)&=
\tilde\omega\bigl( F(\phi_{rel})F(j^0_{rel})\bigr)=\frac{\sqrt{\rho_0}}{4\gamma}
\coth\beta\gamma,
\end{align*}
hence leading to the following equality, as operators:
\begin{equation*}
F(j^0_{rel}) = \frac{\sqrt{\rho_0}}{\gamma} F(\phi_{rel}).
\end{equation*}

From proposition \ref{pr:vir-thm} and \ref{pr:var-rel}, it follows that
\begin{equation*}
\tilde\omega\bigl( F(j_{rel})^2 \bigr) = \frac{c_{rel}}{4\gamma}\coth\beta\gamma,
\end{equation*}
and from proposition \ref{pr:rel-fluc-can-pair}, 
\begin{equation*}
\lim_{\beta\to\infty} c_{rel} = \frac{\rho_0}{\gamma}.
\end{equation*}
Therefore
\begin{equation*}
\tilde\omega^g\bigl( F(j_{rel})^2 \bigr) = \frac{\rho_0}{4\gamma^2} = 
\tilde\omega^g\bigl( F(j^0_{rel})^2 \bigr).
\end{equation*}
This implies necessarily
\begin{equation*}
\tilde\omega^g\Bigl(\bigl[ F(j_{rel})-F(j^0_{rel}) \bigr]^2\Bigr)=0,
\end{equation*}
and again the equality of the operators
\begin{equation*}
F(j_{rel})=F(j^0_{rel})
\end{equation*}
in the ground state, as a result of coarse graining.
\end{proof}

The physical interpretation of this proposition is the following. For non-zero 
temperatures, the relative current consists of two terms. One of them is $j^0_{rel}$, 
which has a non-trivial contribution to the fluctuation of the relative current only 
if $\rho_0 >0$, i.e. whenever the
gauge symmetry is spontaneously broken. The other term contains no more reference to the
zero mode, in other words to the condensate. Therefore it is clear that the fluctuation
operator $F(j^0_{rel})$ contains all the information of the fluctuations of what one could
call the \emph{condensate current}, or the current between the condensates interacting
through the Josephson junction. The important equality
\begin{equation*}
F(j^0_{rel}) = \frac{\sqrt{\rho_0}}{\gamma} F(\phi_{rel})
\end{equation*}
is nothing but a rigorous translation, on the level of the fluctuations, of
the popular statement: ``the (superfluid, condensate) current is the gradient of the
phase''. The second statement of the proposition shows that the quantum
effects on the level of the fluctuations, originating from the spontaneous symmetry
breaking, are only present in the ground state. This of course is popular wisdom,
already experienced in many models \cite{verbeure:1992, michoel:1999b, michoel:2000}, 
but expressed here in a mathematically rigorous fashion for our model.

Finally, it may come as a surprise that our results show no dependence on $\varphi$, the
expectation value for the phase difference between the condensates (see equation
(\ref{eq:phase-diff})). This however is a simple consequence of the description 
of the system in
its mathematically simplest form, using the operators $b^\sharp_{\pm,k}$ (\ref{eq:b-ops}), 
which yield a $\varphi$-independent description of the system. Indeed, from a mathematical 
point of view, the system can not be expected to behave different for different $\varphi$, 
since e.g. the eigenvalues of the Hamiltonian $E^\pm_k$ are $\varphi$-independent.

If one is interested in the physics following from a non-zero $\varphi$, i.e. if one wants
to derive typical Josephson currents proportional to $\sin\varphi$, one needs to
work with the bare operators $a^\sharp_{(1,2),k}$. In particular, consider 
the relative current fluctuation operator defined by
\begin{equation*}
F_L(j_{rel}^\varphi)=\frac{1}{2\gamma V^{1/2}}\sum_k a^*_{1,k}a_{2,k}+a^*_{2,k}a_{1,k}
-\omega(a^*_{1,k}a_{2,k}+a^*_{2,k}a_{1,k}),
\end{equation*}
in stead of (\ref{eq:rel-cur-fluc}).

It can easily be calculated that this operator satisfies
\begin{align*}
\lim_{L\to\infty}\bigl[ F_L(j_{rel}),F_L(j_{rel}^\varphi)\bigr] &=0\\
\lim_{L\to\infty}\bigl[ F_L(n_{rel}),F_L(j_{rel}^\varphi)\bigr] &=ic_{rel}\sin\varphi\\
\delta^\omega\bigl(F_L(j_{rel}^\varphi)\bigr) &= i\sin\varphi F_L(n_{rel}).
\end{align*}
Together with the results above, this establishes that the limiting fluctuation operator
$F(j_{rel}^\varphi)$ is given by
\begin{equation}\label{eq:rel-cur-sinus}
F(j_{rel}^\varphi) = F(j_{rel}) \sin\varphi,
\end{equation}
where (\ref{eq:rel-cur-sinus}) is to be understood in terms of the equivalence between
fluctuation operators (\ref{eq:coarse-grain}), i.e. (\ref{eq:rel-cur-sinus}) follows from
\begin{equation*}
\lim_{L\to\infty}\omega\Bigl( \bigl(F_L(j_{rel}^\varphi)-\sin\varphi F_L(j_{rel})\bigr)^2
\Bigr)=0.
\end{equation*}
The variances of $F(j_{rel}^\phi)$ and its dynamics, computed from proposition 
\ref{pr:dyn} show the explicit $\varphi$-dependence, and its typical `collapse and revival'
properties, found in the experiments \cite{wright:1997}.


\begin{thebibliography}{10}

\bibitem{andersonm:1995}
Anderson M.H., Ensher J.R., Matthews M.R., Wiemann C.E., and Cornell E.A.
\newblock {\em Science}, 269:198, 1995.

\bibitem{davis:1995}
Davis K.B., Mewes M.O., Andrews M.R., van Druten~N.J., Durfee D.S., Kurn D.M.,
  and Ketterle W.
\newblock {\em Physical Review Letters}, 75:2969, 1995.

\bibitem{bradley:1995}
Bradley C.C., Sackett C.A., Tollett J.J., and Hulet R.G.
\newblock {\em Physical Review Letters}, 75:1687, 1995.

\bibitem{mewes:1996}
Mewes M.O., Andrews M.R., van Druten~N.J., Durfee D.S., Kurn D.M., and Ketterle
  W.
\newblock {\em Science}, 273:84, 1996.

\bibitem{nozieres:1990}
Nozieres P. and Pines D.
\newblock {\em The theory of quantum liquids, Vol. 2}.
\newblock Addison-Wesley, Massachusets, 1990.

\bibitem{gardiner:1997}
Gardiner W.
\newblock {\em Physical Review A}, 56:1414, 1997.

\bibitem{fannes:1982}
Fannes M., Pul{\`e} J.V., and Verbeure A.
\newblock On {B}ose condensation.
\newblock {\em Helvetica Physica Acta}, 55:391 -- 399, 1982.

\bibitem{goderis:1989b}
Goderis D. and Vets P.
\newblock Central limit theorem for mixing quantum systems and the
  {CCR}-algebra of fluctuations.
\newblock {\em Communications in Mathematical Physics}, 122:249, 1989.

\bibitem{goderis:1990}
Goderis D., Verbeure A., and Vets P.
\newblock Dynamics of fluctuations for quantum lattice systems.
\newblock {\em Communications in Mathematical Physics}, 128:533 -- 549, 1990.

\bibitem{michoel:1999b}
Michoel T. and Verbeure A.
\newblock {G}oldstone boson normal coordinates in interacting {B}ose gases.
\newblock {\em Journal of Statistical Physics}, 96(5/6):1125 -- 1162, 1999.

\bibitem{wright:1997}
Wright E.M., Wong T., Collet M.J., Tan S.M., and Walls D.F.
\newblock {\em Physical Review A}, 56:591, 1997.

\bibitem{wang:1999}
Wang X.G., Pan S.H., and Yang G.Z.
\newblock {\em Physica A}, 274:484, 1999.

\bibitem{villain:1998}
Villain P. and Lewenstein M.
\newblock Dephasing of josephson oscillations between two coupled bose-einstein
  condensates.
\newblock {\em Physical Review A}, 59:2250, 1998.

\bibitem{michoel:2000}
Michoel T. and Verbeure A.
\newblock {Goldstone Boson} normal coordinates.
\newblock {\em {Preprint-KUL-TF-2000/02} ({arXiv:math-ph/0001033})}, 2000.

\bibitem{goderis:1989}
Goderis D., Verbeure A., and Vets P.
\newblock Non-commutative central limits.
\newblock {\em Probability Theory and Related Fields}, 82:527 -- 544, 1989.

\bibitem{huang:1967}
Huang K.
\newblock {\em Statistical Mechanics}.
\newblock Wiley, London, 1967.

\bibitem{davies:1972}
Davies E.B.
\newblock {\em Communications in Mathematical Physics}, 28:69, 1972.

\bibitem{fannes:1980b}
Fannes M. and Verbeure A.
\newblock The condensed phase of the imperfect bose gas.
\newblock {\em Journal of Mathematical Physics}, 21(7):1809 -- 1818, 1980.

\bibitem{bratteli:1996}
Bratteli O. and Robinson D.W.
\newblock {\em Operator Algebras and Quantum Statistical Mechanics 2}.
\newblock Springer Berlin, Heidelberg, New York, 1996.

\bibitem{fannes:1977}
Fannes M. and Verbeure A.
\newblock Correlation inequalities and equilibrium states.
\newblock {\em Communications in Mathematical Physics}, 55:125 -- 131, 1977.

\bibitem{fannes:1977b}
Fannes M. and Verbeure A.
\newblock Correlation inequalities and equilibrium states. {II}.
\newblock {\em Communications in Mathematical Physics}, 57:165 -- 171, 1977.

\bibitem{verbeure:1992}
Verbeure A. and Zagrebnov V.A.
\newblock Phase transition and algebra of fluctuation operators in an exactly
  soluble model of a quantum anharmonic crystal.
\newblock {\em Journal of Statistical Physics}, 69:329, 1992.

\end{thebibliography}
\end{document}